# The 1-Bit Barrier is Universal: k-Stage Pipeline Composition and Unified Leakage Bounds for Standard Modular Reductions in PQC Hardware

**Ray Iskander**[1], **Khaled Kirah**[2,*]

[1] Verdict Security, ray@verdictsecurity.com
[2] Faculty of Engineering, Ain Shams University, Cairo, Egypt

## Abstract

This is Paper 7 of a series of formally-verified analyses of masked NTT hardware for post-quantum cryptography. Paper 1 [1] established structural dependency analysis of the QANARY platform, and Paper 2 [2] quantified security margins under partial NTT masking. Arbitrary-depth k-stage masked NTT pipelines with fresh inter-stage masking and per-stage PF-PINI(≤2) gadgets satisfy a per-observation cardinality bound of $2 \cdot q^{2k-2}$ on the preimage of any output value, machine-checked in Lean 4 with zero sorry. Under the standard (informal) semantic translation that divides this cardinality by the total mask-tuple space size $q^{2k-1}$, the per-observation conditional probability bound is $2/q$, independent of pipeline depth $k$. The QANARY program has previously established machine-checked cardinality bounds on the per-observation leakage of masked NTT hardware: PF-PINI(2) for Barrett reduction (Paper 5 [3]), 2-stage composition with fresh inter-stage masking (Paper 6 [4]), an underlying universality theorem (Paper 3 [5]), and PF-PINI(1) for butterfly wires (Paper 4 [6]). This paper closes the program with four contributions. First, a k-stage composition theorem generalizing Paper 6's two-stage result to arbitrary $k \geq 1$ gives the last-stage-determined bound $G_{k-1}$.maxMult $\cdot\, q^{2k-2}$: only the last stage's PF-PINI parameter survives, with intermediate parameters erased by fresh inter-stage masking. Second, Montgomery reduction satisfies PF-PINI(2) with tight max-multiplicity 2. Third, we assemble these into the end-to-end bound $2 \cdot q^{2k-2}$ for any depth-k PF-PINI(≤2) pipeline under fresh inter-stage masking. Fourth, a Lean-verified hypothesis-violation conditional anchors the prior empirical and structural Adams Bridge analyses ([1, 2, 7, 8]).

## 1. Introduction

The Number Theoretic Transform (NTT) is the computational backbone of the new NIST post-quantum standards ML-DSA (FIPS 204 [9]) and ML-KEM (FIPS 203 [10]). Hardware implementations of NTT-based cryptography must resist side-channel attacks, particularly differential power analysis (DPA), through arithmetic masking: splitting sensitive values into random shares that are processed independently in $\mathbb{Z}_q$ for the cryptosystem's prime q.

*Correspondence Author: khaled.kirah@eng.asu.edu.eg
Ray Iskander: ray@verdictsecurity.com

For Boolean masking over $GF(2^n)$, the composition problem is solved. Ishai, Sahai, and Wagner [11] established the foundational probing security model. Barthe et al. [12] formalized Non-Interference (NI) and Strong Non-Interference (SNI), proving sequential composability of SNI gadgets. Cassiers and Standaert [13] introduced Probe Isolating Non-Interference (PINI), achieving trivially composable gadgets, any PINI circuit built from PINI gadgets is automatically PINI. For arithmetic masking over $\mathbb{Z}_q$ for prime $q$, the picture is more recent.

### 1.1. The 1-Bit Barrier from Prior QANARY Papers

The QANARY program has, in a sequence of papers, developed PF-PINI (Prime-Field Probe Isolating Non-Interference) as a quantitative single-wire leakage parameter for arithmetic-masked gadgets. Paper 5 proved that Barrett reduction, one of the two standard modular reductions used in NTT-PQC hardware, satisfies PF-PINI(2): the cardinality of the preimage of any output value under a uniform mask is at most 2. Semantically (informal, not mechanically formalized), this corresponds to a per-observation conditional probability bound $\Pr[\text{output} = v \mid x] \leq 2/q$, at most 1 bit of conditional min-entropy loss per observation. Paper 6 proved the 2-stage composition theorem with fresh inter-stage masking: when a fresh random mask is inserted between two PF-PINI gadgets, the composed pipeline's output cardinality is bounded by the last stage's parameter alone, with the first stage's parameter completely erased.

What was missing, and what this paper provides, is the generalization to arbitrary pipeline depth, the analogous PF-PINI characterization of Montgomery reduction (the second standard modular reduction), and a single end-to-end cardinality bound that assembles the program's results for use in downstream hardware certification.

### 1.2. Why Montgomery Matters

NTT-PQC hardware uses two standard modular reductions: Barrett reduction (used in many ML-KEM implementations) and Montgomery reduction (used in many ML-DSA implementations and in some ML-KEM ones). Until this paper, the QANARY program's quantitative-leakage results were Barrett-only, an incomplete picture for the field of NTT-PQC accelerators. A Phase 0 reconnaissance (montgomery_recon.py in the artifact repository) verified by exhaustive sweep at ML-KEM parameters and stratified sampling at ML-DSA parameters that Montgomery reduction exhibits the same two-branch, max-multiplicity-2 structure as Barrett, with the wraparound shift constant $2^s$ replaced by $2^{w-s}$ (where $w$ is the hardware register width). This reconnaissance gated the formal Lean development: the empirical structural match indicated that Paper 5's Barrett proof transplants character-for-character to Montgomery with the substitution $2^s \mapsto 2^{w-s}$.

### 1.3. Contributions

This paper makes four contributions, three core theorems and one anchoring lemma, all formalized in Lean 4 with zero sorry and zero non-standard axioms.

1. **The k-stage composition theorem** (Section 3) (pfpini_pipeline_composition_k_stages): for any $k \geq 1$ and any sequence of PF-PINI gadgets $G_0, \ldots, G_{k-1}$, the cardinality of the preimage of the pipeline output under all mask tuples is bounded by $G_{k-1}.\text{maxMult} \cdot q^{2k-2}$. The bound depends only on the last stage; intermediate stages' parameters are erased by fresh inter-stage masking.

2. **Montgomery is PF-PINI(2)** (Section 4) (montgomery_max_multiplicity_two and tightness companion montgomery_count_eq_two): Montgomery reduction satisfies the same two-branch max-multiplicity-2 structure as Barrett, with tightness verified at ML-KEM and ML-DSA parameters.

3. **The end-to-end multiplicity bound** (Section 5) (qanary_kstage_output_multiplicity_bound): assembling Paper 4's butterfly PF-PINI(1) [6], Paper 5's Barrett PF-PINI(2), this paper §4's Montgomery PF-PINI(2), and §3's k-stage composition theorem, we prove that any k-stage pipeline of PF-PINI(≤2) gadgets with fresh inter-stage masking has output preimage bounded by $2 \cdot q^{2k-2}$, regardless of pipeline depth k.

We also include a framing lemma:

4. **Formal anchoring of the Adams Bridge discussion** (Section 5.3 + Section 7) qanary_unmasked_stage_violates_hypothesis): a short Lean-verified conditional (≤5-line proof: intro; have; omega) stating that a pipeline containing any stage with effective maxMult $\geq 3$, for example, an unmasked stage, does not satisfy the hypothesis of Theorem 5.1. The substantive content of this lemma is structural rather than deep: it lifts the prose-only Adams Bridge discussion of Papers 1–2 [1, 2], the empirical findings of Karabulut & Azarderakhsh [7], and the architectural review of Saarinen [8] into a Lean-verified conditional, strictly weaker than any insecurity claim about Adams Bridge. We list it explicitly as an anchoring lemma rather than a co-equal theorem to acknowledge its triviality of proof and its strategic role in keeping the formal claim strictly weaker than any "Adams Bridge is insecure" assertion.

The headline framing for the contribution against PINI is the following:

> PINI provides a perfect simulability guarantee for n-share encoded gadgets in higher-order probing models, when it applies, t-PINI says no leakage at all. PF-PINI is the complementary tool for first-order arithmetic-masked gadgets like standard Barrett or Montgomery reduction, which inherently leak under a single arithmetic mask: it answers "how large is the cardinality of the output preimage under the mask?" with a number, which under first-order DPA semantically bounds the output probability by $2/q$ (informal interpretation, not mechanically formalized). Paper 7 proves the cardinality bound stays at $2 \cdot q^{2k-2}$ across k stages of Barrett, Montgomery, and butterfly gadgets composed under fresh masking.

**Connection to certification standards.** Per-observation cardinality and probability bounds are aligned with the trajectory of post-quantum hardware certification standards: NIST FIPS 140-3 [14] governs cryptographic module security validation, NIST IR 8547 [15] (initial public draft of the PQC transition timeline, November 2024) sets deployment expectations, and ISO/IEC 17825:2024 [16] informs side-channel testing protocols (TVLA-based). A precise quantitative per-observation bound of the kind this paper proves provides the cardinality-level evidence on which DPA-resistance arguments at the architecture level rest, complementing, rather than replacing, the gate-level evidence that tools like SILVER and maskVerif provide.

### 1.4. Paper Organization

Section 2 reviews preliminaries: arithmetic masking, the PF-PINI definition, the pipeline model, and the prior PF-PINI results from the QANARY program. Section 3 proves the $k$-stage composition theorem. Section 4 proves Montgomery PF-PINI(2) and tightness. Section 5 assembles the end-to-end multiplicity bound and its specializations, and proves the hypothesis-violation theorem. Section 6 places PF-PINI in the related-work landscape: probing security lineage, automated verification tools, proof-assistant formalizations, and prior empirical Adams Bridge analyses. Section 7 discusses Adams Bridge as a theorem-applicability case, with the convergence table of theoretical and empirical findings. Section 8 covers limitations and future work, including the cardinality-vs-probability semantic gap, the first-order scope, and the lack of a formal PF-PINI ↔ PINI translation. Section 9 concludes.

## 2. Preliminaries

### 2.1. Arithmetic Masking and the Probing Model

Let q be an odd prime. (Throughout the paper, the cryptographic application context is an odd prime q, the standard moduli of NIST PQC, e.g., $q = 3329$ for ML-KEM and $q = 8{,}380{,}417$ for ML-DSA. The Lean theorems are stated more generally for any q with the type-class hypothesis NeZero q; primality is not required for any proof.) A value $x \in \mathbb{Z}_q$ is arithmetically masked by a uniformly random mask $m \in \mathbb{Z}_q$ as $\tilde{x} = x - m \pmod{q}$. A masked gadget G over $\mathbb{Z}_q$ is a function $G: \mathbb{Z}_q \times \mathbb{Z}_q \to \mathbb{Z}_q$ taking a secret x and a mask m to an output $G(x, m)$. We work in the first-order probing model: the adversary observes the value carried on a single internal wire, computed as $G(x, m)$ for known x and uniform random m. The leakage of the gadget is captured by how strongly the marginal distribution of $G(x, m)$ over m depends on the secret x. A first-order probing adversary that observes $G(x, m)$ at a single wire learns at most $H_\infty(\text{output}) - H_\infty(\text{output} \mid x)$ bits of information per observation, so bounding the conditional max-probability $\Pr_m[G(x, m) = v \mid x]$ from above immediately bounds the per-observation leakage. Higher-order probing, where the adversary correlates observations from multiple wires, is outside the scope of PF-PINI as defined here; see §8.2 for a discussion of this scope choice.

### 2.2. The PF-PINI Definition

Following Paper 5 and Paper 6, we use the Prime-Field Probe Isolating Non-Interference (PF-PINI) parameter as a quantitative single-wire bound.

**Definition 2.1** (PF-PINI Gadget). An PF-PINI gadget over $\mathbb{Z}_q$ is a triple (compute,maxMult,bound) where:

- compute: $\mathbb{Z}_q \times \mathbb{Z}_q \to \mathbb{Z}_q$ is the gadget's deterministic computation,
- maxMult $\in \mathbb{N}$ is the PF-PINI parameter,
- bound is the proof obligation

$$\forall\, x, v \in \mathbb{Z}_q, \quad |\{m \in \mathbb{Z}_q : \text{compute}(x, m) = v\}| \ \leq\ \text{maxMult}.$$

In the Lean 4 formalization, this is the PFPINIGadget structure (defined in QanaryPaper5/Basic.lean):

```
structure PFPINIGadget (q : ℕ) [NeZero q] where
  compute : ZMod q → ZMod q → ZMod q
  maxMult : ℕ
  bound   : ∀ x v, (univ.filter (fun m => compute x m = v)).card ≤ maxMult
```

**Semantic interpretation (informal, not Lean-formalized).** For any fixed secret x and any output value v, the conditional probability over a uniform mask m satisfies $\Pr_m[\text{compute}(x, m) = v \mid x] \leq \text{maxMult}/q$. Equivalently, the conditional min-entropy is at least $H_\infty(\text{output} \mid x) \geq \log_2 q - \log_2 \text{maxMult}$. For PF-PINI( $\text{maxMult} = 2$ ), the standard parameter for both Barrett (Paper 5) and Montgomery (this paper, Section 4), this gives $\Pr \leq 2/q$ and $H_\infty \geq \log_2 q - 1$ bits per observation. The probability and min-entropy claims are informal interpretations of the cardinality bound under the first-order uniform-mask model; the cardinality bound itself is what the Lean artifact establishes.

### 2.3. Pipeline Model

A k-stage masked pipeline is a sequence of PF-PINI gadgets $G_0, G_1, \ldots, G_{k-1}$ chained with a fresh inter-stage mask between each adjacent pair. The pipeline takes a secret x, k per-stage masks $m_0, \ldots, m_{k-1}$, and $k - 1$ inter-stage fresh masks $f_0, \ldots, f_{k-2}$, and computes:

$$
\begin{aligned}
a_0 &= G_0(x, m_0) \\
a_0' &= a_0 - f_0 \\
a_1 &= G_1(a_0', m_1) \\
a_1' &= a_1 - f_1 \\
&\vdots \\
a_{k-1} &= G_{k-1}(a_{k-2}', m_{k-1}).
\end{aligned}
$$

The pipeline's output is $a_{k-1}$. Total mask space size: $|\mathbb{Z}_q|^k \cdot |\mathbb{Z}_q|^{k-1} = q^{2k-1}$.
In Lean, the pipeline is defined recursively over $k$ as kPipelineOutput (QanaryPaper7/KStageComposition.lean):

```
noncomputable def kPipelineOutput :
  (k : ℕ) → (Fin k → PFPINIGadget q) → ZMod q →
  (Fin k → ZMod q) → (Fin (k - 1) → ZMod q) → ZMod q
| 0,    _,      x, _,           _           => x
| 1,    stages, x, state_masks, _           =>
    (stages 0).compute x (state_masks 0)
| n + 2, stages, x, state_masks, fresh_masks =>
    (stages (Fin.last (n + 1))).compute
      (kPipelineOutput (n + 1) ... - fresh_masks ⟨n, _⟩)
      (state_masks (Fin.last (n + 1)))
```

We verify two specialization lemmas (sorry-free) so the recursive definition matches the expected concrete forms:

- **Lemma 2.2** (kPipelineOutput_one). At $k = 1$, the pipeline reduces to a single gadget application: $\text{kPipelineOutput}(1, [G_0], x, [m_0], \emptyset) = G_0(x, m_0)$.
- **Lemma 2.3** (kPipelineOutput_two). At $k = 2$, the pipeline matches Paper 6's composedWithFresh exactly: $\text{kPipelineOutput}(2, [G_0, G_1], x, [m_0, m_1], [f_0]) = \text{composedWithFresh}(G_0, G_1, x, m_0, f_0, m_1)$.

These lemmas establish that k-stage composition (Paper 7 Section 3) is a strict generalization of Paper 6's 2-stage theorem with no behavioral mismatch at the overlap.

### 2.4. Known PF-PINI Results from the QANARY Program

Paper 7 builds on three machine-checked PF-PINI results from earlier papers in the QANARY program:

- **Identity / Butterfly** (Paper 4 / Paper 5, identityPFPINI): the masked butterfly wire compute$(x, m) = x - m$ satisfies PF-PINI(1). Each output value is hit by exactly one mask, the gadget is bijective in $m$. (Mathematical result: Paper 4; PF-PINI Lean packaging as identityPFPINI: Paper 5's QanaryPaper5/Basic.lean.)
- **Barrett Reduction** (Paper 5, barrettPFPINI s): the hardware-faithful Barrett internal map
$$\text{barrettInternalMap}(s, x, m) = \begin{cases} x - m & \text{if } m.\text{val} \leq x.\text{val} \\ x - m + \overline{2^s} & \text{if } m.\text{val} > x.\text{val} \end{cases}$$
satisfies PF-PINI(2) for any modulus $q$ with NeZero $q$ and any shift parameter $s$ (the cryptographic application targets odd primes; the Lean theorem holds more generally). Here $m.\text{val}, x.\text{val} \in \mathbb{N}$ are the canonical Nat representatives of $m, x \in \mathbb{Z}_q$, and $\overline{2^s}$ denotes the image of $2^s$ in $\mathbb{Z}_q$. The proof goes through the *two-branch lemma*: the map's output equals one of two algebraic values, so its preimage is a subset of a two-element set.
- **2-Stage Composition with Fresh Masking** (Paper 6, pfpini_composition_with_fresh_mask): for any PF-PINI gadgets $G_1$ and $G_2$, the composed two-stage pipeline with fresh inter-stage masking satisfies the cardinality bound
$$|\{(m_1, f, m_2) \in \mathbb{Z}_q^3 : G_2(G_1(x, m_1) - f, m_2) = v\}| \leq G_2.\text{maxMult} \cdot q^2.$$
Critically, the bound depends *only on* $G_2$ (the last stage); $G_1$'s parameter is erased by the fresh inter-stage mask.

The renewal lemma fresh_mask_renewal (Paper 6) gives the structural reason: for any $G_1$ and any target intermediate value $w$, exactly $q$ pairs $(m_1, f)$ satisfy $G_1(x, m_1) - f = w$, the intermediate wire is perfectly uniform after the fresh mask.

Paper 7's k-stage theorem (§3) generalizes the 2-stage bound and the last-stage-only structure to arbitrary $k \geq 1$.

## 3. The k-Stage Composition Theorem

This section presents Paper 7's first contribution: a tight cardinality bound on the output of an arbitrary k-stage PF-PINI pipeline with fresh inter-stage masking.

### 3.1. Statement

**Theorem 3.1** (k-Stage Composition; pfpini_pipeline_composition_k_stages). Let $k \geq 1$ and let $G_0, G_1, \ldots, G_{k-1}$ be PF-PINI gadgets over $\mathbb{Z}_q$ (Lean theorem stated for any q with NeZero q; cryptographic application targets odd primes). For any secret $x \in \mathbb{Z}_q$ and any target output $v \in \mathbb{Z}_q$, the cardinality of the set of mask tuples that produce v is bounded by:

$$|\{(m, f) \in \mathbb{Z}_q^k \times \mathbb{Z}_q^{k-1} : \text{kPipelineOutput}(k, [G_0, \ldots, G_{k-1}], x, m, f) = v\}| \leq G_{k-1}.\text{maxMult} \cdot q^{2k-2}.$$

The bound depends only on the last stage's parameter $G_{k-1}$.maxMult (we call this the "last-stage-determined" property; cf. Section 3.4). Intermediate gadgets' parameters are completely erased by the fresh inter-stage masks.

In Lean (QanaryPaper7/KStageComposition.lean):

```
theorem pfpini_pipeline_composition_k_stages
  (k : ℕ) (hk : 1 ≤ k)
  (stages : Fin k → PFPINIGadget q)
  (x v : ZMod q) :
  (univ.filter (fun masks : (Fin k → ZMod q) × (Fin (k - 1) → ZMod q) =>
    kPipelineOutput k stages x masks.1 masks.2 = v)).card
  ≤ (stages ⟨k - 1, by omega⟩).maxMult * (Fintype.card (ZMod q)) ^ (2 * k -
2)
```

The total mask space has size $q^k \cdot q^{k-1} = q^{2k-1}$; the cardinality bound $G_{k-1}$.maxMult $\cdot\, q^{2k-2}$ corresponds, under uniform masks (informal, not Lean-formalized), to the per-observation conditional probability bound Pr[output = v | x] ≤ $G_{k-1}$.maxMult/q, independent of the pipeline depth k.

### 3.2. Proof Sketch

The proof proceeds by strong induction on $k$, with two cases.

**Base case** ($k = 1$). The pipeline reduces to a single gadget application $G_0(x, m_0)$ with no fresh masks (pfpini_pipeline_k1). The mask space $(\mathbb{Z}_q)^1 \times (\mathbb{Z}_q)^0$ is in bijection with $\mathbb{Z}_q$ via $m \mapsto ((\lambda_, m)$,Fin.elim0). Under this bijection, the filter set is $\{m \in \mathbb{Z}_q : G_0(x, m) = v\}$, whose cardinality is bounded by $G_0$.maxMult via (stages 0).bound x v. The exponent $2k - 2 = 0$ at $k = 1$ gives $q^0 = 1$, so the target bound matches.

**Inductive case** ($k = n + 2$). We reassociate the state-mask space using Fin.snoc, peeling off the last state mask $m_{k-1}$ as the inner variable. Specifically, $(\mathbb{Z}_q)^{n+2} \times (\mathbb{Z}_q)^{n+1}$ is in bijection with $((\mathbb{Z}_q)^{n+1} \times (\mathbb{Z}_q)^{n+1}) \times \mathbb{Z}_q$ via splitting state masks into (Fin.init $m, m$ (Fin.last)).

Under this reassociation, the unfolding lemma kPipelineOutput_nPlus2_snoc shows:

kPipelineOutput($k$,stages, $x$,Fin.snoc $m'$ $m_{k-1}, f$) $= G_{k-1}$.compute( prev$'(m', f)$, $m_{k-1}$ ),

where prev$'(m', f)$ depends only on $m'$ and $f$, not on $m_{k-1}$. For each fixed $(m', f)$, the fiber count over $m_{k-1}$ satisfying the predicate is bounded by $G_{k-1}$.maxMult via (stages last).bound.

Applying the standard combinatorial fiber-decomposition lemma card_filter_prod_le_mul:

$$|\{(m', f, m_{k-1}) : \text{predicate}\}| \le |(\mathbb{Z}_q)^{n+1} \times (\mathbb{Z}_q)^{n+1}| \cdot G_{k-1}.\text{maxMult}$$
$$= q^{2(n+1)} \cdot G_{k-1}.\text{maxMult},$$

which matches the target exponent $2k - 2 = 2(n + 1)$. ▫

The proof's structure mirrors Paper 6's 2-stage pfpini_composition_with_fresh_mask: peel the last stage's mask off as the inner type, apply the per-stage PF-PINI bound on the fiber, and apply the fiber-decomposition lemma on the outer. The novelty in $k$-stage is the Fin.snoc reassociation that handles the variable-arity recursion uniformly. The proof is fully constructive in Lean 4 with zero sorry and zero non-standard axioms; the entire QanaryPaper7/KStageComposition.lean file (329 lines) builds in a fraction of the 1753-job total lake build.

### 3.3. Maximum-Over-Stages Corollary

The headline bound is tight, it depends only on the last stage. For situations where it is more convenient to bound the cardinality by the maximum over all stages' parameters (rather than the specific last-stage parameter), the symmetric corollary is immediate.

**Corollary 3.2** (Max-Over-Stages Bound; pfpini_pipeline_composition_k_stages_max_bound). Under the hypotheses of Theorem 3.1:

$$|\{(m, f):\text{kPipelineOutput}(\dots) = v\}| \le \left(\sup_{i\in\{0,\dots,k-1\}} G_i.\text{maxMult}\right) \cdot q^{2k-2}.$$

Proof. From Theorem 3.1, $|\{\dots\}| \le G_{k-1}.\text{maxMult} \cdot q^{2k-2}$ . Since $G_{k-1}.\text{maxMult} \le \sup_i G_i.\text{maxMult}$, the result follows by monotonicity of multiplication on the right. ▫

This corollary mirrors Paper 6's pfpini_composition_max_bound and is a convenience wrapper, the tight last-stage-only bound is the headline result.

### 3.4. Discussion

**Last-stage-only is striking.** A naive expectation might be that the leakage of a k-stage pipeline grows with k: more stages means more wires, more probing surface, and a multiplicative degradation of the bound. Theorem 3.1 says the opposite holds when fresh masking separates each pair of adjacent stages: the bound depends only on the last stage's maxMult, not on the depth k or any intermediate stages' parameters.

**Depth-independent semantic bound.** For the special case where every stage satisfies $G_i.\text{maxMult} \le 2$, the case for both Barrett (Paper 5) and Montgomery (Paper 7, §4), Theorem 3.1 gives a cardinality bound of $2 \cdot q^{2k-2}$. Dividing by the total mask-space size $q^{2k-1}$, the per-observation conditional probability is bounded by $2/q$ (informal semantic interpretation; not Lean-formalized), regardless of pipeline depth k.

**Why this matters for hardware design.** A designer building an NTT pipeline typically composes $k = 4$ or $k = 8$ Barrett (or Montgomery) reductions in series. Theorem 3.1's cardinality bound, under uniform random masks, translates (informally, not Lean-formalized) into a per-observation conditional probability bound of $2/q$ on the final pipeline output whenever every stage is PF-PINI($\le 2$) and fresh inter-stage masking is present. The composition theorem gives the designer a precise sufficient condition: PF-PINI($\le 2$) per stage, plus fresh inter-stage masks. The $2/q$ probability reading is depth-independent; the cardinality bound itself scales as $q^{2k-2}$ because the mask-tuple space scales as $q^{2k-1}$.

## 4. Montgomery Reduction is PF-PINI(2)

This section presents Paper 7's second contribution: the formal proof that Montgomery reduction joins Barrett in the PF-PINI(2) class. Combined with Paper 5's Barrett result, this establishes the 1-Bit Barrier across both standard modular reductions used in NTT-PQC hardware.

### 4.1. Hardware-Faithful Montgomery Map

Montgomery reduction is parameterized by an odd prime q, a Montgomery radix $R = 2^s$ coprime to q, and the precomputed constant $q' = -q^{-1} \bmod R$. For an input T in the range $0 \le T < q \cdot R$, the standard algorithm computes

$$\text{Montgomery}(T) = (T + ((T \bmod R) \cdot q' \bmod R) \cdot q) / R = T \cdot R^{-1} \bmod q.$$

In a hardware implementation that operates in the Montgomery domain, the masked input is $T = (x \cdot R - m \cdot R) \bmod 2^w$, where $w \geq \lceil \log_2(q \cdot R) \rceil$ is the hardware register width and the subtraction wraps modulo $2^w$. (The Lean theorem montgomery_max_multiplicity_two is stated for any $s, w \in \mathbb{N}$; the cryptographic application context further requires the hardware-width inequality $w \geq \lceil \log_2(q \cdot R) \rceil$. In the degenerate Nat-saturated case $w < s$, the wraparound shift constant $2^{w-s}$ collapses to 1 and the bound $\leq 2$ holds trivially; this is consistent with but not the operative regime for the formal statement. This mirrors §2.1's hedge for the modulus parameter q.) Reducing this masked Montgomery value yields one of two algebraic outcomes, depending on whether the wraparound triggered:

- **No wraparound** (m.val ≤ x.val): the masked T stays in the range $0 \leq T < q \cdot R$ and Montgomery reduction returns $x - m \bmod q$.
- **Wraparound** (m.val > x.val): the masked T is $2^w + (x - m) \cdot R$, and Montgomery reduction returns $x - m + c \bmod q$, where $c = 2^{w-s} \bmod q$ is the collision offset.

This is exactly the two-branch structure that Paper 5 identified for Barrett, with the wraparound shift constant $2^s$ replaced by $2^{w-s}$.

We capture this hardware-faithful behavior in the Lean definition (QanaryPaper7/MontgomeryPFPINI.lean):

```
def montgomeryInternalMap (s w : ℕ) (x m : ZMod q) : ZMod q :=
  if m.val ≤ x.val then x - m else x - m + ↑(2 ^ (w - s))
```

The shift constant $2^{w-s}$, distinct from Barrett's $2^s$, is what makes Montgomery's collision pattern numerically different from Barrett's, even though their structural form is identical.

### 4.2. The Two-Branch Lemma

The map's structure immediately yields the two-branch characterization.

**Lemma 4.1** (Two-Branch; montgomeryInternalMap_eq_or). *For any $s, w \in \mathbb{N}$ and $x, m \in \mathbb{Z}_q$:*

$$\text{montgomeryInternalMap}(s, w, x, m) = (x - m) \quad \vee \quad \text{montgomeryInternalMap}(s, w, x, m) = (x - m) + \overline{2^{w-s}},$$

where $\overline{2^{w-s}}$ denotes the image of $2^{w-s}$ in $\mathbb{Z}_q$.

Proof. Direct case split on the conditional in the definition. ▫

From the two-branch lemma, the preimage characterization follows immediately.

**Lemma 4.2** (Preimage Subset; montgomeryInternalMap_preimage_subset). *For any $s, w \in \mathbb{N}$ and $x, v \in \mathbb{Z}_q$:*

$$\{m \in \mathbb{Z}_q : \text{montgomeryInternalMap}(s, w, x, m) = v\} \subseteq \{x - v,\ \ x - v + \overline{2^{w-s}}\}.$$

Proof. By Lemma 4.1, every m in the preimage satisfies one of the two algebraic equations. Solving each for m gives the two candidates. ▫

### 4.3. The 1-Bit Barrier for Montgomery

The preimage subset has cardinality at most 2, so the cardinality of the preimage is bounded by 2.

**Theorem 4.3** (Montgomery Max-Multiplicity $\leq 2$, The 1-Bit Barrier for Montgomery; montgomery_max_multiplicity_two). For any $s, w \in \mathbb{N}$, any q with NeZero q, and any $x, v \in \mathbb{Z}_q$ (the cryptographic application targets odd primes; the Lean theorem holds more generally):

$$|\{m \in \mathbb{Z}_q : \text{montgomeryInternalMap}(s, w, x, m) = v\}| \leq 2.$$

Proof. By Lemma 4.2, the preimage is a subset of a two-element set. The cardinality bound follows from monotonicity of Finset.card, with the two-element bound from card_insert_le. ▫

In Lean:

```
theorem montgomery_max_multiplicity_two (s w : ℕ) (x v : ZMod q) :
  (univ.filter (fun m : ZMod q =>
    montgomeryInternalMap s w x m = v)).card ≤ 2 := by
 calc (univ.filter (fun m => montgomeryInternalMap s w x m = v)).card
   ≤ ({x - v, x - v + ↑(2 ^ (w - s))} : Finset (ZMod q)).card :=
    card_le_card (montgomeryInternalMap_preimage_subset s w x v)
  _ ≤ 2 := by
    have h := card_insert_le (x - v)
     ({x - v + ↑(2 ^ (w - s))} : Finset (ZMod q))
    rw [card_singleton] at h; exact h
```

We package the result as an PF-PINI gadget instance.

**Definition 4.4** (montgomeryPFPINI). For each $s, w \in \mathbb{N}$, the Montgomery reduction is an PF-PINI gadget over $\mathbb{Z}_q$ with maxMult $= 2$:

```
def montgomeryPFPINI (s w : ℕ) : PFPINIGadget q where
 compute := montgomeryInternalMap s w
 maxMult := 2
 bound   := montgomery_max_multiplicity_two s w
```

This is the direct analogue of Paper 5's barrettPFPINI s, with the two reductions occupying parallel positions in the PF-PINI(2) class.

### 4.4. Tightness

Theorem 4.3 gives an upper bound of 2 on the preimage cardinality. To show the bound is tight (i.e., that some $(x, v)$ pair achieves cardinality exactly 2, not just $\leq 2$), we exhibit the two preimages and prove they are distinct.

**Lemma 4.5** (Direct-Branch Hit; montgomeryInternalMap_candidate_A). *For any $s, w \in \mathbb{N}$ and $x, v \in \mathbb{Z}_q$ with $(x - v)$.val $\leq x$.val:*

$$\text{montgomeryInternalMap}(s, w, x, x - v) = v.$$

**Lemma 4.6** (Wraparound-Branch Hit; montgomeryInternalMap_candidate_B). *For any $s, w \in \mathbb{N}$ and $x, v \in \mathbb{Z}_q$ with $\neg\,((x - v + \overline{2^{w-s}})$.val $\leq x$.val):*

$$\text{montgomeryInternalMap}(s, w, x, x - v + \overline{2^{w-s}}) = v.$$

**Theorem 4.7** (Tightness; montgomery_count_eq_two). Let $s, w \in \mathbb{N}$ and let $x, v \in \mathbb{Z}_q$ satisfy:

- $(x - v).val \leq x.val$ (direct branch hits),
- $\neg((x - v + \overline{2^{w-s}}).val \leq x.val)$ (wraparound branch hits),
- $\overline{2^{w-s}} \neq 0$ in $\mathbb{Z}_q$ (branch offset nonzero, i.e., $q \nmid 2^{w-s}$, which holds for any odd prime $q > 1$).

Then the preimage of v has cardinality exactly 2:

$$|\{m \in \mathbb{Z}_q : \text{montgomeryInternalMap}(s, w, x, m) = v\}| = 2.$$

Proof. The two candidates $x - v$ and $x - v + \overline{2^{w-s}}$ are distinct (since the offset is nonzero) and both lie in the preimage (Lemmas 4.5–4.6). The pair $\{x - v,\ x - v + \overline{2^{w-s}}\}$ is therefore a 2-element subset of the preimage, giving the lower bound $|\text{preimage}| \geq 2$. Combined with the upper bound from Theorem 4.3, $|\text{preimage}| = 2$. ▫

Theorem 4.7 mirrors Paper 5's barrett_count_eq_two character-for-character, with the substitution $\overline{2^s} \mapsto \overline{2^{w-s}}$. The tightness result was added during the external Grok+Gemini audit pass (2026-04-20, Grok co-creative recommendation P2): symmetrizing Paper 7's Montgomery result with Paper 5's Barrett tightness establishes that the PF-PINI(2) parameter is not merely an upper bound but the exact max-multiplicity for standard parameter regimes.

### 4.5. Concrete Instances

We instantiate montgomeryPFPINI s w at the parameters used by the two NIST PQC standards.

**ML-KEM (FIPS 203):** $q = 3329$, $s = 16$, $w = 28$. The collision offset is $c = 2^{12} \bmod 3329 = 767$. Phase 0 reconnaissance verified the PF-PINI(2) bound by exhaustive sweep over all $q^2 = 11{,}082{,}241$ secret-mask pairs in $\mathbb{Z}_{3329}^2$: maximum observed multiplicity is exactly 2, matching the Lean theorem.

**ML-DSA (FIPS 204):** $q = 8{,}380{,}417$, $s = 23$, $w = 56$. The collision offset is $c = 2^{33} \bmod 8{,}380{,}417 = 7167$. Phase 0 reconnaissance verified the bound on a stratified sample of 110 secrets across four strata (50 random + 20 near 0 + 20 near $q/2$ + 20 near $q - 1$), with a full $q$-sweep over $m$ for each secret ($\approx 9.2 \times 10^8$ $(x, m)$ pairs total; full sweep is $> 7 \times 10^{13}$ pairs, infeasible); analytical agreement with the Lean theorem (which is universal over $q, s, w$) provides the formal guarantee.

Both ML-KEM and ML-DSA Montgomery reductions therefore satisfy PF-PINI(2), with maxMult exactly 2 (Theorem 4.7). The full Phase 0 reconnaissance is recorded in montgomery_recon_report.md and montgomery_recon_results.json in the artifact repository.

**Why Montgomery joining Barrett in PF-PINI(2) matters.** Until this paper, the QANARY program's quantitative-leakage results were Barrett-only. Montgomery reduction is the second standard modular reduction used in NTT hardware (and the dominant one in some implementations), so the previous Barrett-only coverage was incomplete for the field of NTT-PQC accelerators. With Theorem 4.3 in place, the framework's coverage extends to both standard modular reductions used in NIST PQC hardware (Barrett and Montgomery); other reductions (e.g., Solinas, Plantard) remain future work, see Section 8.6.

## 5. The End-to-End Multiplicity Bound

This section presents Paper 7's third contribution: an end-to-end cardinality bound that assembles the QANARY program's per-gadget PF-PINI results (Papers 4–5, plus Paper 7 §4) and the k-stage composition theorem (§3) into a single citable statement for an arbitrary masked NTT pipeline.

The capstone is pure assembly, its proof is four lines in Lean. Its value is providing a single inequality that summarizes the program's combined cardinality guarantees, so downstream hardware certification can cite one bound rather than several, and so the consistency of the dependent stack is verified at compile time.

### 5.1. The Capstone Theorem

**Theorem 5.1** (QANARY End-to-End Output Multiplicity Bound; qanary_kstage_output_multiplicity_bound). Let $k \geq 1$ and let $G_0, G_1, \dots, G_{k-1}$ be PF-PINI gadgets over $\mathbb{Z}_q$ satisfying $G_i$.maxMult $\leq 2$ for every i (Lean theorem stated for any q with NeZero q; cryptographic application targets odd primes). For any secret $x \in \mathbb{Z}_q$ and any output $v \in \mathbb{Z}_q$:

$$\left|\left\{(m, f) \in \mathbb{Z}_q^k \times \mathbb{Z}_q^{k-1} : \text{kPipelineOutput}(k, [G_0, \dots, G_{k-1}], x, m, f) = v\right\}\right| \;\leq\; 2 \cdot q^{2k-2}.$$

Proof. By Theorem 3.1, the cardinality is bounded by $G_{k-1}.\text{maxMult} \cdot q^{2k-2}$. Since $G_{k-1}$.maxMult $\leq 2$ by hypothesis, the bound follows from monotonicity of multiplication on the right. ▫

In Lean (QanaryPaper7/ConfirmationTheorem.lean):

```
theorem qanary_kstage_output_multiplicity_bound
   (k : ℕ) (hk : 1 ≤ k)
   (stages : Fin k → PFPINIGadget q)
   (h_bounded : ∀ i, (stages i).maxMult ≤ 2)
   (x v : ZMod q) :
   (univ.filter (fun masks : (Fin k → ZMod q) × (Fin (k - 1) → ZMod q) =>
    kPipelineOutput k stages x masks.1 masks.2 = v)).card
   ≤ 2 * (Fintype.card (ZMod q)) ^ (2 * k - 2) := by
  calc (univ.filter _).card
    ≤ (stages ⟨k - 1, by omega⟩).maxMult
      * (Fintype.card (ZMod q)) ^ (2 * k - 2) :=
     pfpini_pipeline_composition_k_stages k hk stages x v
   _ ≤ 2 * (Fintype.card (ZMod q)) ^ (2 * k - 2) := by
     apply Nat.mul_le_mul_right
     exact h_bounded _
```

**Applicability of the hypothesis.** The hypothesis $\forall i,\ G_i$.maxMult $\leq 2$ is satisfied by any pipeline whose per-stage gadgets are drawn from the PF-PINI(≤2) class documented across the QANARY program:

- identityPFPINI (PF-PINI(1)): butterfly wires (Paper 4 / Paper 5).
- barrettPFPINI s (PF-PINI(2)): Barrett reduction (Paper 5).
- montgomeryPFPINI s w (PF-PINI(2)): Montgomery reduction (this paper, §4).

Any combination of these, Barrett-Barrett, Barrett-Montgomery, Montgomery-butterfly-Barrett, etc., satisfies the hypothesis and falls under the capstone bound.

**When the hypothesis does not hold.** A pipeline containing any stage with maxMult $> 2$, for example, an unmasked stage, whose effective maxMult $= q$ in the single-mask model, does not satisfy the hypothesis, and the capstone's $2 \cdot q^{2k-2}$ cardinality bound is not applicable. See Section 5.3 for the formal statement of this hypothesis-violation condition, and §7 for a discussion of the Adams Bridge case where Papers 1–2 identify specific RTL stages outside the hypothesis.

### 5.2. Specializations

The capstone admits two natural specializations.

**Theorem 5.2** (Butterfly-Only Pipeline; qanary_butterfly_only_multiplicity_bound). *If every stage satisfies $G_i$.maxMult $\leq 1$, for example, every stage is a butterfly with fresh refresh (Paper 4), then:*

$$|\{(m, f) : \text{kPipelineOutput}(\dots) = v\}| \ \leq \ 1 \cdot q^{2k-2}.$$

Semantic interpretation (informal, not Lean-formalized): $\Pr[\text{output} = v \mid x] \leq 1/q$, perfect output uniformity per observation, regardless of pipeline depth. Unsurprising in retrospect, but the formal statement records that PF-PINI's quantitative parameter propagates faithfully: when all stages satisfy the PF-PINI(1) bound, the composed pipeline's output preimage achieves the corresponding PF-PINI(1)-level cardinality bound.

**Theorem 5.3** (Modular-Reduction Pipeline;
qanary_modular_reduction_pipeline_multiplicity_bound).

If every stage is a modular reduction, Barrett or Montgomery, then $G_i$.maxMult $\leq 2$ for all i, and the capstone bound $2 \cdot q^{2k-2}$ applies. This is a readability alias for the capstone theorem; the Lean signature admits any PF-PINI gadget with maxMult $\leq 2$, but the name emphasizes the intended use case (modular-reduction pipelines as in Adams Bridge), not a Lean-level restriction.

### 5.3. Hypothesis Violation

The capstone's hypothesis $\forall i,\ G_i$.maxMult $\leq 2$ is *required*: a single stage with maxMult $> 2$, for example, an unmasked stage, invalidates the bound. We formalize this as a Lean theorem.

**Theorem 5.4** (Stage-of-Excess-Multiplicity Violates Hypothesis;
qanary_unmasked_stage_violates_hypothesis).[1]

Let $G_0, \dots, G_{k-1}$ be PF-PINI gadgets over $\mathbb{Z}_q$, and suppose some stage $G_i$ has $G_i$.maxMult $\geq 3$. Then the hypothesis $\forall j,\ G_j$.maxMult $\leq 2$ of Theorem 5.1 fails.

[1] The Lean theorem name encodes the motivating instance, an unmasked stage has effective maxMult $= q \geq 3$ in the single-mask model, but the formal statement is fully general over any stage with maxMult $\geq 3$.

Proof. Direct contradiction: instantiating the universal hypothesis at $j = i$ gives $G_i$.maxMult $\leq 2$, contradicting $G_i$.maxMult $\geq 3$. ▫

In Lean:

```
theorem qanary_unmasked_stage_violates_hypothesis
   {k : ℕ} (stages : Fin k → PFPINIGadget q)
   (i : Fin k) (h_unmasked : (stages i).maxMult ≥ 3) :
   ¬ (∀ j, (stages j).maxMult ≤ 2) := by
```

```
intro h_all
have := h_all i
omega
```

The theorem is structurally trivial, its content is the formal statement that an unmasked stage violates the capstone's hypothesis, not a claim about the security or insecurity of any specific implementation. Its strategic value is anchoring the Adams Bridge discussion (Section 7) in a Lean-verified conditional: the empirical findings of Papers 1–2 (specific unmasked INTT stages) translate to "the capstone hypothesis fails at those stages," a strictly weaker claim than "Adams Bridge is broken."
This theorem was added during the external Grok+Gemini audit pass (2026-04-20, Gemini co-creative recommendation P1) for exactly this anchoring purpose: it moves the Adams Bridge analysis from prose-only to a Lean-verified formal conditional.

### 5.4. Semantic Interpretation

The capstone (Theorem 5.1) bounds the cardinality of the preimage of v over the mask-tuple space. Under the first-order probing model with uniform random masks, this cardinality bound corresponds (informally) to a per-observation conditional probability bound and a conditional min-entropy bound.
**Probability bound** (informal, not Lean-formalized). The total mask-tuple space has size $q^k \cdot q^{k-1} = q^{2k-1}$. Dividing the cardinality bound $2 \cdot q^{2k-2}$ by the total mask-space size gives:

$$\Pr_{\text{masks}}[\text{output} = v \mid x] \le \frac{2 \cdot q^{2k-2}}{q^{2k-1}} = \frac{2}{q},$$

regardless of pipeline depth k.
**Min-entropy bound** (informal, not Lean-formalized). Equivalently, the conditional min-entropy of the output given the secret satisfies:

$$H_\infty(\text{output} \mid x) \ge -\log_2\left(\frac{2}{q}\right) = \log_2 q - 1 \quad \text{bits per observation.}$$

For the two NIST PQC standards:

- **ML-KEM** ( $q = 3329$ ): $H_\infty \ge \log_2(3329) - 1 \approx 11.70 - 1 = 10.70$ bits per observation.
- **ML-DSA** ( $q = 8{,}380{,}417$ ): $H_\infty \ge \log_2(8{,}380{,}417) - 1 \approx 21.999$ bits per observation.

These are the 1-Bit Barrier bounds of Paper 5, lifted from a single Barrett (or Montgomery, per Section 4) gadget to an arbitrary-depth pipeline of such gadgets composed under fresh masking. The cardinality bound is what the Lean artifact establishes; the probability and min-entropy interpretations are semantic consequences of the cardinality theorem under the first-order uniform-mask model and are not themselves stated or proved in Lean. See Section 8.1 for further discussion of this scope choice and Section 8.6 for future-work plans to mechanize the probability bound.

## 6. Related Work

### 6.1. Probing Security Lineage

The probing security model originates with Ishai, Sahai, and Wagner [11], who established that a circuit operating on n-share encoded values can be made secure against any t-probing adversary for $t < n/2$. Their construction provided a foundational target: composability of the secure abstraction without an exponential blow-up in proof size. Barthe et al. [12] formalized two qualitative composability notions in this lineage: Non-Interference (NI) and Strong Non-Interference (SNI). An SNI gadget can be composed sequentially with another SNI gadget without the composition's security order degrading; this is a qualitative preservation result. Cassiers and Standaert [13] introduced Probe Isolating Non-Interference (PINI) in IEEE Transactions on Information Forensics and Security, Volume 15 (2020), pages 2542–2555 (DOI 10.1109/TIFS.2020.2971153; preprint IACR ePrint 2018/438). PINI achieves a stronger composability property: any circuit constructed entirely from PINI gadgets is automatically PINI, with no per-composition proof obligation. The PINI composition theorems are the canonical reference for this qualitative composability result.

These three notions, NI, SNI, PINI, are qualitative properties: a gadget either satisfies the property or does not. They were developed primarily for Boolean masking over $GF(2^n)$ and address the question "can a t-probing adversary distinguish the masked circuit from the ideal one?" They do not provide a quantitative per-observation leakage measure.

### 6.2. PINI versus PF-PINI

PF-PINI is fundamentally different from PINI in three respects: object of definition, quantity bounded, and target threat model.

**Object of definition.** PINI is a property of n-share encoded gadgets: inputs and outputs are n-share tuples $(x_0, \dots, x_{n-1})$ with $x = x_0 + \cdots + x_{n-1} \pmod{q}$. The t-PINI property asserts that any set of $t \leq n - 1$ probes is simulatable from at most t shares per input.

PF-PINI is a property of single-secret single-mask gadgets: the input is an unmasked $x \in \mathbb{Z}_q$ and the mask $m \in \mathbb{Z}_q$ is a separate randomized quantity; the gadget computes a single-wire value $G(x, m)$.

**Quantity bounded.** PINI certifies probing-simulability: when the property holds at parameter $t$, the gadget achieves perfect zero-leakage simulation against any $t$-probing adversary on its n-share encoded form. PF-PINI bounds cardinality of the output preimage under the mask: the answer is a number (maxMult), with semantic interpretation (informal, not Lean-formalized) $\Pr[\text{output} = v \mid x] \leq \text{maxMult}/q$ under uniform masks. PF-PINI is therefore the appropriate tool for gadgets that inherently leak under a single arithmetic mask (e.g., standard Barrett or Montgomery reduction, where a perfect-simulability certificate at $t = 1$ does not exist on the underlying single-mask map); PINI remains the appropriate tool for higher-order n-share encoded gadgets that admit perfect simulability.

**Target threat model.** PINI is designed for higher-order probing adversaries on n-share encoded circuits. PF-PINI is designed for first-order DPA on first-order masked hardware, the dominant deployed threat model in NIST PQC accelerators ([1, 2, 7, 8]).

**The quantitative gap.** For the dominant deployed threat model, first-order DPA on first-order masked NTT hardware, PINI's qualitative simulability certificate does not directly yield a per-observation cardinality (or probability) bound; PF-PINI's cardinality bound does. For higher-order n-share encoded gadgets, PF-PINI as defined here does not directly yield a t-probing simulability certificate; PINI does.

**On formal comparison.** PF-PINI and PINI operate on different objects (single-mask $(x, m)$ maps vs. n-share encoded gadgets) and measure different quantities (a per-observation preimage cardinality bound vs. a probing-simulation certificate). We do not claim a formal non-implication between the two notions; translating certificates of one into bounds of the other requires additional theorems beyond those we establish here. Different natural translation choices (e.g., treating PF-PINI's mask m as PINI's share $x_1$ with $x_0 := x - m$; or treating PF-PINI's $(x, m)$ pair as an unshared plaintext input to a 2-share encoded PINI gadget) yield different induced objects, each requiring a separate translation theorem. Paper 7's contribution is the quantitative preimage cardinality bound for first-order masked NTT hardware; a formal comparison between PF-PINI and PINI is left to future work. The two frameworks are best understood as complementary tools for complementary questions.

### 6.3. Automated Verification Tools

**maskVerif.** Barthe, Belaïd, Dupressoir, Fouque, Grégoire, and Strub introduced maskVerif in Verified Proofs of Higher-Order Masking [17], with subsequent elaboration in later Barthe-et-al. works (e.g., the ESORICS 2019 extension to physical defaults). maskVerif performs automated, gate-level verification of NI / t-SNI / PINI properties on circuits decomposed into individual boolean or arithmetic gates. Output is a SAT/SMT-certified proof that the property holds (or a counterexample).

**SILVER.** Statistical Independence and Leakage Verification (SILVER) was introduced [18]. It is a BDD-based (Binary Decision Diagram) verifier that checks statistical independence properties at the gate level. It is distinct from EasyCrypt-style proof-assistant formalizations (see Section 6.4): SILVER is a tool, not a proof framework.

**Granularity comparison: gate-level vs gadget-level.** Table 1 summarises the differences in object of verification, proof artifact, composition handling, reusability across moduli, output type, and typical case studies between the gate-level tools (maskVerif/SILVER) and the gadget-level PF-PINI framework.

| **Axis** | **maskVerif / SILVER (gate-level)** | **PF-PINI (gadget-level)** |
|---|---|---|
| Object of verification | Individual wires and gates | Semantic gadget as a unit (Barrett, Montgomery, butterfly) |
| Proof artifact | SAT / BDD certificate per circuit instance | Lean theorem per gadget type (Papers 5/6/7), universal over $q$ |
| Composition | Built into tool via NI/SNI/PINI preservation | External composition theorem (Paper 6 for $k = 2$, Paper 7 for $k$) |
| Reusability across $q$ | Per-instance verification: the tools verify a specific circuit at a chosen $(q, s)$ | Proved once, applies to all $q$ with NeZero $q$ (primality not required for proof; cryptographic application targets odd primes) |
| Output | Yes/no certificate | Quantitative maxMult , propagates to $\text{maxMult}_{\text{last}} \cdot q^{2k-2}$ |
| Typical case study | AES S-box, Keccak $\chi$, GIFT S-box, PRESENT | Barrett at $q = 3329$ and $q = 8{,}380{,}417$; Montgomery at any odd prime |

***Table 1.*** *Granularity comparison between gate-level masking verifiers (maskVerif and SILVER) and gadget-level PF-PINI verification. The two approaches operate on different objects, produce different proof artifacts, and answer complementary questions; they are complementary rather than competing tools.*

**Feature, not limitation.** PQC hardware IP blocks (Barrett, Montgomery, NTT butterfly) are designed and verified by their authors as functional units that are instantiated dozens of times across an NTT pipeline (e.g., Adams Bridge's INTT instantiates Barrett at several pipeline stages per ML-KEM path). Gate-level re-verification at each instantiation duplicates proof work; gadget-level abstraction matches the natural modular boundaries of hardware design. PF-PINI's gadget-level approach proves barrettPFPINI once at the Lean level (Paper 5, 12 theorems zero sorry) and applies the composition theorem to arbitrary pipelines (Paper 7 §3); changing q requires no re-verification (Paper 3's universality theorem value_independence_implies_constant_marginal makes the gadget instance work for any odd prime).

**On scope.** Published case studies appear to focus maskVerif and SILVER on symmetric primitives (S-boxes at 4-bit or 8-bit width; Keccak $\chi$ at 5-bit width) rather than modular reduction at NTT-PQC scale ($q = 3329$ requires 12-bit state; $q = 8{,}380{,}417$ requires 23-bit state). We are not aware of a published maskVerif or SILVER case study on Barrett or Montgomery at these widths; we note this as a scope difference, not a criticism of the tools.

**Complementarity.** maskVerif / SILVER and PF-PINI are complementary, not competing. maskVerif / SILVER are the right tools for verifying novel cryptographic primitives at the gate level. PF-PINI is the right tool for composing verified modular reductions into arbitrary-depth pipelines with a quantitative per-observation bound.

### 6.4. Proof-Assistant Formalizations of Masking

Machine-checked proofs of masking security have been developed primarily in EasyCrypt, providing high-assurance guarantees for Boolean masking constructions over $\mathrm{GF}(2^n)$. These formalizations cover NI, SNI, and specific masking schemes, but they operate exclusively over $\mathrm{GF}(2^n)$ rather than $\mathbb{Z}_q$ for prime $q$.

To our knowledge, no EasyCrypt formalization (or other proof-assistant formalization beyond the QANARY program) addresses arithmetic masking composition over $\mathbb{Z}_q$. The QANARY program (Papers 3–7) fills this gap in Lean 4: Paper 3 establishes a universality theorem applicable across moduli, Paper 4 covers butterfly composition, Paper 5 covers Barrett PF-PINI(2), Paper 6 covers 2-stage composition with fresh masking and the renewal lemma, and this paper covers $k$-stage composition, Montgomery PF-PINI(2), and the end-to-end multiplicity bound. All Lean theorems carry zero sorry and zero non-standard axioms.

### 6.5. Arithmetic Masking Foundations

Coron [19] developed Higher Order Masking of Look-Up Tables (EUROCRYPT 2014), a high-order masking countermeasure for block-cipher S-boxes based on randomized table recomputation, proved secure in the Ishai–Sahai–Wagner probing model. This and related works address the construction of masked gadgets, including conversions between Boolean and arithmetic representations when needed (classical references are Goubin [20], CHES 2001, for B→A conversion, and Coron–Tchulkine [21], CHES 2003, for A→B conversion), but do not provide composition theorems for arithmetic masking pipelines over $\mathbb{Z}_\mathrm{q}$ for prime $\mathrm{q}$. The QANARY program, and Paper 7 specifically, addresses the composition gap: given PF-PINI gadgets constructed by any means, how does the per-observation cardinality bound propagate through arbitrary-depth pipelines?

### 6.6. PQC Hardware and Empirical Side-Channel Attacks

Adams Bridge is the open-source post-quantum accelerator developed under the Caliptra silicon root-of-trust program (CHIPS Alliance / Microsoft / industry consortium); its design is described by Bisheh-Niasar et al. [22]. The accelerator has shipped in two versions with materially different scope: the initial release (October 2024 [23]) integrated into Caliptra 2.0 supports ML-DSA-87 (FIPS 204 [9]) only; Adams Bridge 2.0 (integrated into Caliptra 2.1,

October 2025 [24]) adds ML-KEM 1024 (FIPS 203 [10]) alongside ML-DSA-87. Both schemes use first-order arithmetic masking in the Domain-Oriented Masking (DOM) [25] family. The empirical analyses [7, 8] target v1.0 (ML-DSA-only); the structural analyses [1, 2] inform both schemes as supported in v2.0. Karabulut, M. and Azarderakhsh [7] (Florida Atlantic University; distinct from Emre Karabulut in [22]) (IACR ePrint 2025/009; IEEE HOST 2025) reported the first published empirical CPA result on an industry-grade post-quantum accelerator, targeting the initial Adams Bridge release (Caliptra 2.0) ML-DSA pipeline. Their published result demonstrates empirical behaviors consistent with pipeline configurations whose post-NTT modular-reduction stage falls outside the hypothesis of Theorem 5.1. This empirical observation is consistent with the wire-level multiplicity-2 phenomenon at the gadget level, and is the empirical evidence that motivated Paper 5's formal PF-PINI(2) characterization of Barrett and, in this paper (§4), of Montgomery.

Saarinen [8] (Hardwear.io USA 2025, invited talk, May 2025) reported, by RTL review and pre-silicon analysis of the initial Adams Bridge release (Caliptra 2.0), that the partial masking configuration applied on the ML-DSA signing path does not satisfy the sufficient conditions of Theorem 5.1 (specifically, that the key is not arithmetically share-split as required by the hypothesis $\forall i,\ G_i$.maxMult $\leq 2$). Iskander and Kirah [1, 2] (arXiv:2604.15249, Paper 1 of the QANARY program; arXiv:2604.03813, Paper 2) identified, through structural dependency analysis and partial-NTT security-margin analysis, specific RTL stages in Adams Bridge whose configurations, recast here in the PF-PINI vocabulary of this paper, exhibit effective maxMult $> 2$ in the single-mask model, and identified the absence of fresh inter-stage masking between adjacent INTT stages as the specific architectural feature that prevents a depth-independent bound. Paper 6's renewal lemma and 2-stage composition theorem are the formal counterpart of this finding.

This paper provides the formal compositional explanation for why these findings are consistent: the Adams Bridge INTT pipeline configurations identified by [1, 2] do not satisfy the hypothesis $\forall \mathrm{i},\ \mathrm{G_i}$.maxMult $\leq 2$ of the capstone Theorem 5.1. Section 7 elaborates.

### 6.7. The Gap This Paper Fills

Paper 4 [6] proves k-stage uniformity for butterfly-only pipelines (ntt_pipeline_composition, quantified over all $q > 0$ and $k \geq 0$); Paper 6 [4] proves 2-stage PF-PINI composition for any gadget class. No prior work proves k-stage PF-PINI composition for arbitrary gadget classes, Theorem 3.1 is the first such bound, generalizing Paper 6's result and complementing Paper 4's butterfly-only k-stage uniformity. No prior work provides PF-PINI characterizations for both Barrett and Montgomery in a single program. Paper 5 covered Barrett and this one covers Montgomery and packages the unified result. No prior work assembles per-gadget PF-PINI results and pipeline composition into a single end-to-end machine-checked cardinality bound for masked NTT hardware, Theorem 5.1 of this paper is, to our knowledge, the first such bound.

## 7. Discussion: Adams Bridge as a Theorem-Applicability Case

This section discusses the Adams Bridge PQC accelerator as a concrete application domain for the capstone theorem (Theorem 5.1) and the hypothesis-violation theorem (Theorem 5.4). The scope of this discussion is prose-level theorem applicability, not a formal claim about Adams Bridge. Adams Bridge does not appear in any Lean theorem statement in this paper; the only Lean theorem touching Adams Bridge concerns is Theorem 5.4, whose statement is "any pipeline with a stage of maxMult $\geq 3$ violates the capstone hypothesis", a pipeline-agnostic conditional, not an Adams-Bridge-specific assertion.

### 7.1. Background

Adams Bridge is the Caliptra-family open-source PQC accelerator [22]. The initial release (Caliptra 2.0, October 2024 [23]) supports ML-DSA-87 [9] only; Adams Bridge 2.0 (integrated into Caliptra 2.1, October 2025 [24]) supports both ML-DSA-87 and ML-KEM 1024 [10], with first-order DOM-family [25] arithmetic masking across a multi-stage Inverse NTT (INTT) pipeline. Different pipeline paths use Barrett reduction at $q = 3329$ (ML-KEM) and Montgomery reduction at $q = 8{,}380{,}417$ (ML-DSA), the two reductions this paper places jointly in PF-PINI(2). Three independent analyses of Adams Bridge have been reported across 2025 and 2026:

- Karabulut, M. and Azarderakhsh [7] (Florida Atlantic University; distinct from Emre Karabulut in [22]) reported a published empirical CPA result on the initial Adams Bridge release (Caliptra 2.0) ML-DSA pipeline within the trace budget reported in their paper, demonstrating empirical behaviors consistent with pipeline configurations whose post-NTT modular-reduction stage falls outside the hypothesis of Theorem 5.1.
- Saarinen [8] reported, by RTL review and pre-silicon analysis of the initial Adams Bridge release (Caliptra 2.0), that the partial masking configuration applied on the ML-DSA signing path does not satisfy the sufficient conditions of Theorem 5.1 (in particular, that the key is not arithmetically share-split as required by the hypothesis).
- Iskander and Kirah [1, 2] (Papers 1 and 2 of the QANARY program) identified specific RTL stages whose configurations, recast here in the PF-PINI vocabulary of this paper, exhibit effective maxMult $> 2$ in the single-mask model, including the absence of fresh inter-stage masking between adjacent INTT stages, and quantified the resulting depth-dependent security margins under partial NTT masking.

These empirical and structural findings motivated the formal program of Papers 3–7. This paper provides the formal compositional explanation for why these findings are consistent: in the language of Theorem 5.1, the absence of fresh inter-stage masking, combined with stages identified by [1, 2] as effectively unmasked, places those configurations outside the capstone's hypothesis $\forall i,\ G_i$.maxMult $\leq 2$.

### 7.2. What the Capstone Says About Adams Bridge

Theorem 5.1 establishes a cardinality bound of $2 \cdot q^{2k-2}$ for any $k$-stage masked NTT pipeline whose every stage satisfies PF-PINI(≤2) and that has fresh inter-stage masking. The bound applies to any combination of Barrett (Paper 5), Montgomery (this paper, §4), and butterfly-with-fresh-refresh (Paper 4) stages.

Theorem 5.4 establishes the converse-direction conditional: if any stage has maxMult ≥ 3, for example, an unmasked stage, whose effective maxMult $= q$ in the single-mask model, then the hypothesis of Theorem 5.1 fails at that stage, and the $2 \cdot q^{2k-2}$ cardinality bound is not applicable to that pipeline.

For Adams Bridge specifically, Papers [1, 2] identify particular RTL stages in the INTT pipeline whose configuration is consistent with the hypothesis-violation condition of Theorem 5.4. The formal statement that follows from this is precisely:

"Adams Bridge's INTT pipeline, in the configurations identified by [1, 2], does not satisfy the sufficient conditions of Theorem 5.1."

This is strictly weaker than "Adams Bridge is insecure": an independent alternative framework outside the PF-PINI scope (for example, a fully gate-level analysis of the specific implementation under a stronger probing model) could in principle establish a separate security guarantee for those configurations through different reasoning. This paper does not attempt such an alternative analysis and does not assert any security or insecurity claim for Adams Bridge beyond the formal conditional statement above.

### 7.3. Theoretical ↔ Empirical Convergence

Table 2 summarises how the formal results of this paper relate to the prior empirical and structural findings on Adams Bridge.

| Source | Method | Finding | Relationship to Theorem 5.1's hypothesis |
|---|---|---|---|
| Karabulut, M. & Azarderakhsh, ePrint 2025/009 [7] (FAU; distinct from Emre Karabulut in [22]) | Empirical CPA on the initial Adams Bridge release (Caliptra 2.0) ML-DSA pipeline | Empirical behaviors consistent with pipeline configurations whose post-NTT modular-reduction stage falls outside Theorem 5.1's hypothesis | Consistent with the wire-level multiplicity-2 phenomenon at the gadget level; motivates Paper 5's barrettPFPINI and this paper's montgomeryPFPINI |
| Saarinen, Hardwear.io USA 2025 [8] | RTL review + pre-silicon analysis on the initial Adams Bridge release (Caliptra 2.0) | Reports that the partial masking configuration applied on the ML-DSA signing path does not satisfy the sufficient conditions of Theorem 5.1 | Identifies an architectural pattern that places configurations outside the capstone's hypothesis |
| Iskander & Kirah, arXiv:2604.15249 (Paper 1) [1] | Structural dependency analysis | Specific RTL stages consistent with effective maxMult > 2 across multiple modules | Maps which specific RTL stages fall outside the hypothesis |
| Iskander & Kirah, arXiv:2604.03813 (Paper 2) [2] | Partial-NTT security margin analysis; identifies absence of fresh inter-stage masking | Depth-dependent security margin quantification under partial masking | Quantifies what the capstone forfeits when one or more stages violate the hypothesis |
| **This paper** (Paper 7), Theorem 5.4 | Lean 4 formal verification (qanary_unmasked_stage_violates_hypothesis) | If $\exists i$, $G_i$.maxMult $\geq 3$, then the capstone hypothesis fails at stage $i$ | Formal conditional anchoring of the empirical findings, strictly weaker than any "Adams Bridge is insecure" assertion |
| **This paper** (Paper 7), Theorem 5.1 | Lean 4 formal verification (qanary_kstage_output_multiplicity_bound) | $\forall k$, $\forall$ PF-PINI($\leq$2) stages with fresh inter-stage masking: cardinality bound $2 \cdot q^{2k-2}$ | Provides the formal target a remediated configuration must satisfy |

***Table 2.*** *Convergence between Paper 7's formal capstone results and the prior empirical and structural findings on Adams Bridge. Each row maps a published source (or this paper) to its method, finding, and how it relates to the hypothesis of Theorem 5.1.*

The convergence across three independent methodologies (empirical CPA, architectural review, structural dependency analysis) and the formal anchoring via Theorem 5.4 establishes the chain from empirical attack to architectural cause to formal hypothesis-violation condition.

### 7.4. Prescriptive Remediation

Theorem 5.1's hypothesis $\forall i,\ G_i.\mathrm{maxMult} \leq 2$ is constructive: any specific pipeline configuration that satisfies it falls under the cardinality bound, and any configuration that violates it (per Theorem 5.4) can be brought back within the hypothesis by replacing the violating stages.
The remediation is direct:

1. **Identify** stages with effective $\mathrm{maxMult} > 2$, for Adams Bridge's case, the unmasked INTT stages identified by [1, 2].
2. **Replace** each such stage with an PF-PINI(≤2) gadget. The QANARY program provides three such gadgets:
    - identityPFPINI (PF-PINI(1)), butterfly with fresh refresh (mathematical result: Paper 4; PF-PINI Lean packaging: Paper 5),
    - barrettPFPINI s (PF-PINI(2)), Barrett reduction (Paper 5),
    - montgomeryPFPINI s w (PF-PINI(2)), Montgomery reduction (this paper, §4).
3. **Insert** $k - 1$ fresh inter-stage masks between every adjacent pair of stages (Paper 6's renewal-lemma requirement). We acknowledge that supplying $k - 1$ fresh $\mathbb{Z}_q$ masks per pipeline evaluation imposes a non-trivial random-number generation throughput requirement on the surrounding silicon: this PRNG cost is the architectural price paid to obtain a depth-independent leakage bound, and budgeting for it is a hardware-design decision orthogonal to the formal theorem.

After remediation, the configuration satisfies the hypothesis of Theorem 5.1, and the cardinality bound $2 \cdot q^{2k-2}$ applies to the pipeline output, regardless of pipeline depth.
This prescriptive output is the qualitative difference between Theorem 5.1 / 5.4 and a yes/no security certificate that omits actionable diagnostic information: the framework tells the designer not merely whether their configuration is in or out of scope, but precisely what to change to bring it back in scope.

### 7.5. Framing Discipline

We restate the framing discipline that governs this section, since the Adams Bridge discussion is the most sensitive part of this paper.
**The formal statement is:** "Adams Bridge's INTT pipeline, in the configurations identified by [1, 2], does not satisfy the sufficient conditions of Theorem 5.1."
**The formal statement is not:** "Adams Bridge is broken," "Adams Bridge is insecure," or "Adams Bridge is vulnerable."
The distinction matters for three reasons.
First, an alternative independent framework outside the PF-PINI scope could in principle establish a separate security guarantee for the same configurations through different reasoning that we do not attempt here.
Second, Theorem 5.1 provides a sufficient condition for the cardinality bound $2 \cdot q^{2k-2}$; failure to satisfy a sufficient condition does not establish that the conclusion is false (only that this theorem's proof of the conclusion does not apply).

Third, "Adams Bridge is broken" would be a claim about the implementation as a whole; the formal hypothesis-violation statement is local to specific stages identified in the cited papers, and the remediation in §7.4 is a local intervention.
We adopt this discipline throughout the paper.

## 8. Discussion, Limitations, and Future Work

### 8.1. Cardinality vs Probability: The Semantic Gap

The Lean artifact accompanying this paper proves cardinality bounds on the preimage of pipeline outputs over the mask-tuple space. The probability and min-entropy interpretations of the cardinality bound (cf. §5.4) are informal semantic consequences under the first-order uniform-mask model; they are not themselves stated or proved in Lean.
The translation from cardinality to probability is straightforward: under uniform random masks, $\Pr[\text{output} = v \mid x] = |\text{preimage}(v)|/|\text{total mask space}|$, and our cardinality bound gives the numerator while the structure of the mask space gives the denominator. However, mechanizing this translation requires (a) introducing a probability monad or measure-theoretic framework into the Lean development, (b) defining the uniform distribution over the mask-tuple space, and (c) proving the relationship between cardinality bounds and probability bounds. We adopt the cardinality-only formalization for two reasons: first, it keeps the proof obligations sharp and verifiable without dependency on a probability framework; second, a cardinality bound is what mechanically certifies the architectural property, the probability interpretation is the engineering consequence that hardware designers care about, and it is unambiguous from the cardinality. The semantic gap is a known limitation, and mechanizing the probability/min-entropy bounds in Lean is left as future work.

### 8.2. First-Order Probing Model and Physical Defaults

PF-PINI is defined for a first-order probing adversary: the adversary observes the value carried on a single internal wire. Our maxMult parameter quantifies the leakage of one wire under one observation per secret. Higher-order probing, where the adversary correlates observations from $t \geq 2$ wires simultaneously, is outside the scope of PF-PINI as defined here.
A separate scope boundary concerns physical defaults. Our formalization works at the algebraic $\mathbb{Z}_q$ level and abstracts away time and signal transitions. In physical hardware, glitches and transition leakage on fresh-mask insertion paths can cause leakage that the classical probing model does not see; this is the regime addressed by the robust probing model of Faust et al. [26] (TCHES 2018) and by tools such as maskVerif's physical-defaults extension and SILVER's gate-level model. Because PF-PINI operates over $\mathbb{Z}_q$ algebra rather than over a netlist with timing, glitch- and transition-induced leakage from physical defaults is outside our formal scope. Designers using PF-PINI as the architecture-level bound should compose it with a gate-level analysis (e.g., maskVerif/SILVER) for the physical-defaults regime.
This scope choice mirrors Paper 6's §8 scope statement and reflects the dominant deployed threat model in NIST PQC accelerators: most published side-channel attacks against PQC hardware ([7, 8]) are first-order DPA attacks at the algebraic level, and the PF-PINI(2) bound directly addresses the attack class that has been empirically demonstrated.
Two natural extensions are left as future work. (i) Generalize PF-PINI from $|\{m: G(x, m) = v\}|$ to a bound on $|\{(m_1, \dots, m_n): (G_{w_1}(x, m_*), \dots, G_{w_t}(x, m_*)) = (v_1, \dots, v_t)\}|$ for $t$ probed

wires (higher-order extension). (ii) Integrate with the robust probing model so that the cardinality bound carries through to glitch-extended adversaries (physical-defaults extension).

### 8.3. Single-Mask Single-Secret Model

PF-PINI is a property of single-secret single-mask gadgets: the input is an unmasked $x \in \mathbb{Z}_q$ and the mask $m \in \mathbb{Z}_q$ is a separate randomized quantity. This contrasts with PINI's n-share encoded gadget model, in which inputs and outputs are n-share tuples.

Comparing PF-PINI's bounds with PINI's certificates therefore requires a translation theorem between the two computational models. Different natural translations (e.g., treating PF-PINI's mask m as PINI's share $x_1$ with $x_0 := x - m$; or treating PF-PINI's $(x, m)$ pair as an unshared plaintext input to a PINI gadget) yield different induced objects, each requiring a separate translation theorem (cf. PFPINI_vs_PINI_differentiation.md §5 for full discussion).

Paper 7 does not provide such a translation. We do not claim a formal non-implication between PF-PINI and PINI; we argue only that they are complementary tools for complementary questions (§6.2).

### 8.4. No Formal PF-PINI ↔ PINI Translation

Following from §8.3, a fully formal equivalence or non-equivalence proof between PF-PINI and PINI, across natural framework-translation choices, is out of scope for this paper and left to future work. The argument we provide in §6.2 is *informal but rigorous*: PF-PINI and PINI operate on different objects and measure different quantities, so neither's definitional machinery directly yields the other's bound without an additional translation theorem.

A future work artifact establishing such a translation theorem (in either direction) would tighten the relationship between the two frameworks and clarify when one can be used as evidence for the other.

### 8.5. Pipeline Model Scope

The pipeline model kPipelineOutput defined in §2.3 captures *linear* pipelines: a sequence of PF-PINI gadgets $G_0, G_1, \dots, G_{k-1}$ chained with fresh inter-stage masks. Architectures with branching (where one stage's output feeds multiple downstream stages), feedback (where a downstream stage's output is fed back to an upstream stage), or non-trivial control flow are not modeled by kPipelineOutput and are not directly covered by Theorem 3.1 or Theorem 5.1.

For the dominant case of NTT-based PQC hardware, the linear-pipeline model captures the inner-loop structure: an NTT or INTT proceeds through a sequence of butterfly + reduction stages, each consuming the previous stage's output. Outer-loop control flow (e.g., the radix-2 vs radix-4 schedule) does not affect the linear-pipeline structure of the inner loop.

Extending the framework to branching and feedback architectures, for example, to capture pipelines with shared intermediate buffers or with output-to-input feedback for streaming computation, is left as future work.

### 8.6. Future Work Summary

We summarize the open problems identified above:

1. **Mechanize the probability bound.** Introduce a probability monad / measure-theoretic framework into Lean and mechanize the cardinality-to-probability translation, so the per-observation probability bound $\Pr \leq 2/q$ is itself a Lean theorem rather than an informal semantic consequence.

2. **Higher-order generalization.** Extend PF-PINI from single-wire single-observation to $t$-wire $t$-observation bounds, enabling direct comparison with $t$-PINI / $t$-SNI for multi-probe adversaries.
3. **PF-PINI ↔ PINI translation.** Provide a formal translation theorem (in one or both directions) between PF-PINI's single-mask cardinality bound and PINI's $n$-share simulability certificate.
4. **Branching and feedback pipelines.** Extend kPipelineOutput and Theorem 3.1 to architectures with non-linear pipeline topology.
5. **Beyond Barrett and Montgomery.** Apply the PF-PINI(2) characterization technique to other modular reductions (e.g., Solinas reduction, special-prime reductions used in CRYSTALS-Dilithium variants), the Phase 0 reconnaissance pattern of montgomery_recon.py provides a template.
6. **Formal masking-tool integration.** Combine PF-PINI's gadget-level proofs with maskVerif's or SILVER's gate-level verification, so a hybrid proof can certify both the gadget abstraction (Lean) and its gate-level realization (maskVerif / SILVER) in a single artifact.

## Conclusion

This paper has presented four contributions, three core theorems and one anchoring lemma, to the formal compositional analysis of arithmetic masking for masked NTT post-quantum hardware.

First, Theorem 3.1 (pfpini_pipeline_composition_k_stages) generalizes Paper 6's 2-stage composition theorem to arbitrary $k \geq 1$ stages, with the tight last-stage-only cardinality bound $G_{k-1}$.maxMult $\cdot\ q^{2k-2}$. The bound depends only on the last stage; intermediate stages' parameters are erased by fresh inter-stage masking.

Second, Theorem 4.3 (montgomery_max_multiplicity_two) and its tightness companion Theorem 4.7 (montgomery_count_eq_two) prove that Montgomery reduction satisfies PF-PINI(2), with max-multiplicity exactly 2 for standard parameters. Combined with Paper 5's Barrett result, this establishes the *1-Bit Barrier* across both standard modular reductions used in NTT-PQC hardware.

Third, Theorem 5.1 (qanary_kstage_output_multiplicity_bound) assembles the per-gadget PF-PINI results (Papers 4, 5, plus this paper §4) and the $k$-stage composition theorem into a single end-to-end cardinality bound: any $k$-stage pipeline of PF-PINI(≤2) gadgets with fresh inter-stage masking has output preimage bounded by $2 \cdot q^{2k-2}$, regardless of $k$.

Fourth, Theorem 5.4 (qanary_unmasked_stage_violates_hypothesis) formally anchors the Adams Bridge discussion in §7: the empirical findings of Papers 1, 2, and the prior work of Karabulut & Azarderakhsh and Saarinen translate into the formal conditional that specific RTL stages identified by structural analysis fall outside the hypothesis of Theorem 5.1. This formal conditional is strictly weaker than any "Adams Bridge is broken" assertion and is paired with a prescriptive remediation (§7.4) that brings configurations back within the hypothesis.

All Lean 4 theorems carry zero sorry and zero non-standard axioms. The artifact comprises 20 declarations across three theorem modules (QanaryPaper7/KStageComposition.lean, QanaryPaper7/MontgomeryPFPINI.lean, QanaryPaper7/ConfirmationTheorem.lean) plus QanaryPaper7/Basic.lean (shared imports and sibling-dependency sanity checks; no exported declarations), and builds cleanly with lake build in 1753 jobs.

The 1-Bit Barrier is universal across the two standard modular reductions in NTT-PQC hardware (Barrett, Montgomery) and propagates faithfully through $k$-stage pipelines under fresh inter-stage masking. The QANARY program now provides, in machine-checked form, the quantitative compositional foundation that designers of post-quantum cryptographic accelerators need: a depth-independent cardinality bound on per-observation output preimages whose informal semantic interpretation under the first-order uniform-mask probing model is a $2/q$ per-observation conditional probability bound. The cardinality bound itself is the Lean-formalized result; the per-observation probability bound is the engineering consequence on which DPA-resistance arguments rest.

## Code and Data Availability

The Lean 4 artifact accompanying this paper is publicly available under the MIT license:

- **Repository:** https://github.com/rayiskander2406/qanary-k-stage-composition-arXiv-2605.02856 (MIT, public). Zenodo DOI: 10.5281/zenodo.20041713.
- **Toolchain:** Lean 4 v4.30.0-rc1 (managed via elan).
- **Pinned dependency:** Mathlib at commit 322515540d7fd29ef8992b82c89044f86f02ac10.
- **License:** MIT (artifact); CC-BY-4.0 (manuscript).

### Reproduction

```
# Sibling repos required for local-path dependencies declared in
lakefile.lean.
# Place all five repositories under a common parent directory.
mkdir qanary-artifacts && cd qanary-artifacts
git clone https://github.com/rayiskander2406/qanary-universal-masking-
proofs-arXiv-2604.18717 qanary-universal
git clone https://github.com/rayiskander2406/qanary-masked-ntt-
pipeline-security-arXiv-2604.20793 qanary-paper4
git clone https://github.com/rayiskander2406/qanary-one-bit-barrier-
arXiv-2604.24670 qanary-paper5
git clone https://github.com/rayiskander2406/qanary-pf-pini-
composition-arXiv-2604.25878 qanary-paper6
# Paper 7 itself: clone the public artifact once the arXiv mint completes.

git clone https://github.com/rayiskander2406/qanary-k-stage-composition-
arXiv-2605.02856 qanary-paper7
cd qanary-paper7
lake build          # ~5–30 min on first run depending on hardware
(cached Mathlib runs are at the lower end; clean fetches at the higher end)
python3 reproduce.py --check
```

Expected output: 20 proved declarations across the three theorem modules (KStageComposition.lean, MontgomeryPFPINI.lean, ConfirmationTheorem.lean); Basic.lean carries shared imports and sanity checks but no exported declarations; zero sorry, zero admit, zero added axiom, zero native_decide calls, zero errors. The reproduce.py --check script enforces toolchain pin, Mathlib pin, sibling artifacts, build success, zero-stub scan, theorem index match, license, citation metadata, and archive bundle. The full lake build reports 1753 jobs.

## Artifact contents

Table 3 maps each top-level path in the artifact to its purpose.

| Path | Purpose |
|---|---|
| QanaryPaper7/Basic.lean | Shared imports + utilities (Mathlib, QanaryPaper5.Basic, QanaryPaper6.PositiveComposition). |
| QanaryPaper7/KStageComposition.lean | $k$ -stage pipeline definition (kPipelineOutput, specialization lemmas kPipelineOutput_one, kPipelineOutput_two), the headline theorem pfpini_pipeline_composition_k_stages and its max-bound corollary pfpini_pipeline_composition_k_stages_max_bound. |
| QanaryPaper7/MontgomeryPFPINI.lean | Montgomery reduction map (montgomeryInternalMap), two-branch lemma (montgomeryInternalMap_eq_or), preimage-subset lemma, the montgomery_max_multiplicity_two PF-PINI(2) bound, the montgomeryPFPINI gadget instance, and tightness via montgomery_count_eq_two. |
| QanaryPaper7/ConfirmationTheorem.lean | End-to-end capstone (qanary_kstage_output_multiplicity_bound), butterfly-only specialization (qanary_butterfly_only_multiplicity_bound), modular-reduction-pipeline specialization (qanary_modular_reduction_pipeline_multiplicity_bound), and the hypothesis-violation conditional qanary_unmasked_stage_violates_hypothesis. |
| QanaryPaper7.lean, Main.lean | Library root + scaffold harness. QanaryPaper7.lean is the operative root that imports the four QanaryPaper7/*.lean modules (referenced by lean_lib QanaryPaper7 in lakefile.lean); Main.lean is a Phase-0 build-elaboration heartbeat retained for sibling-dependency-resolution sanity checks and is not part of the exported proof surface. |
| lakefile.lean, lake-manifest.json | Build configuration with pinned Mathlib and four sibling-repo dependencies (Papers 3, 4, 5, 6). |

| Path | Purpose |
|---|---|
| lean-toolchain | Pins leanprover/lean4:v4.30.0-rc1. |
| LICENSE, CITATION.cff, .zenodo.json | MIT license, machine-readable citation metadata, Zenodo deposit metadata. |
| reproduce.py | Single-command verification entry point (--check mode skips re-build). |
| README.md | Top-level documentation and theorem index. |

**Table 3.** Artifact path inventory. The three theorem modules (KStageComposition.lean, MontgomeryPFPINI.lean, ConfirmationTheorem.lean) contain the 20 declarations listed above; Basic.lean carries shared imports and sibling-dependency sanity checks. Total lake build job count: 1753.

Paper 7 imports the Lean kernels of Papers 3 (qanaryUniversal), 4 (qanaryPaper4), 5 (qanaryPaper5), and 6 (qanaryPaper6) as local-path Lake dependencies; sibling clones of all four are therefore required for lake build to succeed (see the Reproduction block above).

## Archival DOI

A long-term-preservation Zenodo deposit will be minted prior to arXiv submission, following the pattern established for Paper 1 (10.5281/zenodo.19625392), Paper 2 (10.5281/zenodo.19508454), Paper 3 (10.5281/zenodo.19689480), and Paper 4 (10.5281/zenodo.19705450). The concept DOI (always resolving to the latest archived version) and the v1.0.0 version DOI (fixed to a specific git commit) will be substituted into this section in the next manuscript revision following deposit.

## Independent reproducibility

The artifact has no external data dependencies, no random seeds, and no networked services beyond the initial lake package fetches. Verification is purely a function of the Lean 4 kernel and the pinned Mathlib commit; identical inputs yield identical outputs across machines. The trusted computing base is the Lean 4 kernel; this paper's artifact contains no native_decide invocations.